\def   \ni {\noindent}

\def   \ssk {\vskip  5truept}

\def   \bsk {\vskip 15truept}
 
\def   \newpage {\vfill\eject}
\def   \newline {\hfil\break}

\documentstyle[epsfig]{article}
\begin{document}

\hsize 5truein
\vsize 8truein
\font\abstract=cmr8
\font\keywords=cmr8
\font\caption=cmr8
\font\references=cmr8
\font\text=cmr10
\font\affiliation=cmssi10
\font\author=cmss10
\font\mc=cmss8
\font\title=cmssbx10 scaled\magstep2
\font\alcit=cmti7 scaled\magstephalf
\font\alcin=cmr6 
\font\ita=cmti8
\font\mma=cmr8
\def\ref{\par\noindent\hangindent 15pt}
\null

\title{\ni Long term BeppoSAX-WFC monitoring of GRS~1758-258}                                               

\bsk \bsk
\author{\ni M.~Cocchi$^{1}$, M.J.~Smith$^{2,3}$, J.M.~Muller$^{2,3}$, 
 A.~Bazzano$^{1}$, L.~Natalucci$^{1}$, P.~Ubertini$^{1}$, J.~Heise $^{2}$,   
 and J.J.M.~in~'t~Zand$^{2}$}                                                       
\bsk
\affiliation{ 1) Istituto di Astrofisica Spaziale (IAS/CNR ), via Fosso del Cavaliere, 00133 Roma, Italy.}

\affiliation{ 2) Space Research Organisation Netherlands (SRON), Sorbonnelaan 2, 3584 CA, Utrecht, the Netherlands.}

\affiliation{ 3) {\em BeppoSAX} Science Data Centre, Nuova Telespazio, via Corcolle 19, 00131 Roma, Italy.}

\bsk
\baselineskip = 12pt

\abstract{ABSTRACT\\
\ni
Following a series of deep observations of the Galactic Bulge region performed with the Wide Field Cameras 
telescopes on board {\em BeppoSAX}, it has been possible to study throughout a long time period a number of 
persistent X-ray sources.\\
We report on the behaviour of GRS 1758-258, an X-ray binary suggested as a black hole candidate on the 
basis of its timing and spectral characteristics.  The source has been monitored for almost 2 years since 
August 1996; preliminary results of the data analysis are shown hereafter.
}                                                    
\bsk
\baselineskip = 12pt
\keywords{\ni KEYWORDS: binaries: close --- stars: individual (GRS 1758-258) --- X-rays: stars
}               

\bsk
\baselineskip = 12pt


\text{\ni 1. INTRODUCTION \\
GRS 1758-258, apart from 1E1740.7-2942, is the only persistent hard X-ray / soft Gamma-ray (E$>$75 keV) source in the Galactic Centre 
region (Vargas et al. 1996). 
Like 1E1740.7-2942, the radio counterpart of GRS 1758-258 has relativistic jets (Mirabel et al. 1992) but, despite the high 
radio source location accuracy, an optical counterpart has not been identified yet (Mereghetti et al. 1992).

The source was discovered by the ART-P and SIGMA telescopes on board {\em GRANAT} (Sunyaev et al. 1991) in the spring of 1990.   
The spectrum in the band 30-300 keV was best fitted by a power law model with photon index 
$\Gamma=-1.8 \pm 0.3$.  No strong soft component was observed, suggesting the source was in low (hard) state. \\
GRS 1758-258 was also detected at lower energies (3-30 keV) by {\em SPACELAB-2} and TTM (Skinner 1991) and {\em EXOSAT}-ME.  
Power law spectra with photon index $\Gamma = -1.70 \div -1.75$ were observed. \\
Simultaneous {\em ROSAT} and SIGMA observations in spring 1993 indicated the presence of a soft (E $< 2$ keV) excess 
(Mereghetti et al. 1994), but further analysis of the same data did not confirm the detection of such a soft component 
(Grebenev et al. 1997). \newpage
More recently, a 0.5-10 keV observations with {\em ASCA} (Mereghetti et al. 1997) did not reveal any
strong low-energy excess. The source, observed in March 1995, was found in a low state, and a value of $\Gamma= -1.70 \pm 0.03$
was measured. The fitted value of the column density was ${\em N}_{\rm H} = (1.55 \pm 0.03)\times 10^{22} {\rm cm}^{-2}$, 
supporting a source distance close to the Galactic Centre and ruling out the hypothesis of a massive companion star.

Hard X-ray / soft Gamma ray long term monitoring of the source was performed in 1990-1992 by   
SIGMA (Gilfanov et al. 1993) and ART-P (Grebenev et al. 1997) on board {\em GRANAT}, and by CGRO/BATSE in 1991-1995 (Zhang et al. 1997).
From all the available data it is apparent that the source is a persistent emitter showing strong timing and spectral 
variability. Flux variability of a factor of up to $\sim 10$ was observed (e.g. the dip observed in 1991 both by {\em GRANAT} 
and by BATSE).  The broad band ($\sim30 \div 300$ keV) spectrum of GRS 1758-258 is tipically best fitted by a Sunyaev-Titarchuck 
comptonization model with kT $\sim 30 \div 40$ and $\tau_{disk} \sim 1$, but evidence of spectral transition to pure power 
law emission with softer spectral index was reported (e.g. Grebenev et al. 1997).
Such characteristics, similar to those of Cyg~X-1, lead GRS 1758-258 to be classified as a black hole candidate.
\ssk
\ni

\begin{figure} [hbt]
\centerline{\psfig{file=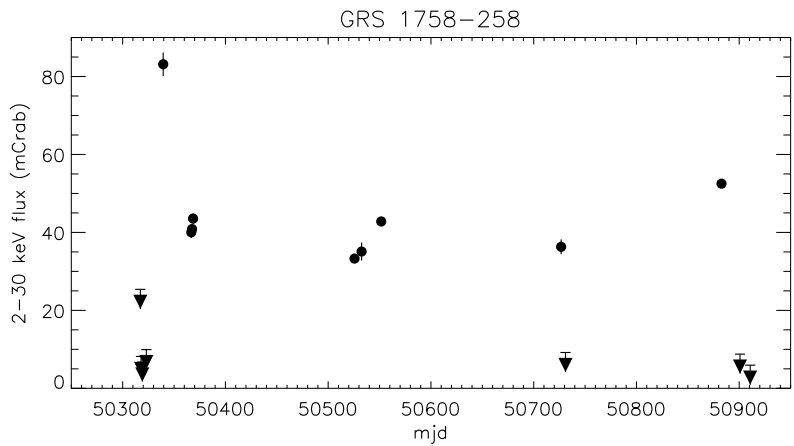, width=11.5cm}}
\caption{FIGURE 1. The 2-30 keV time history of the observed source flux.  Arrows indicate $3\sigma$ upper limits.
}
\end{figure}
\begin{figure} [hbt]
\centerline{\psfig{file=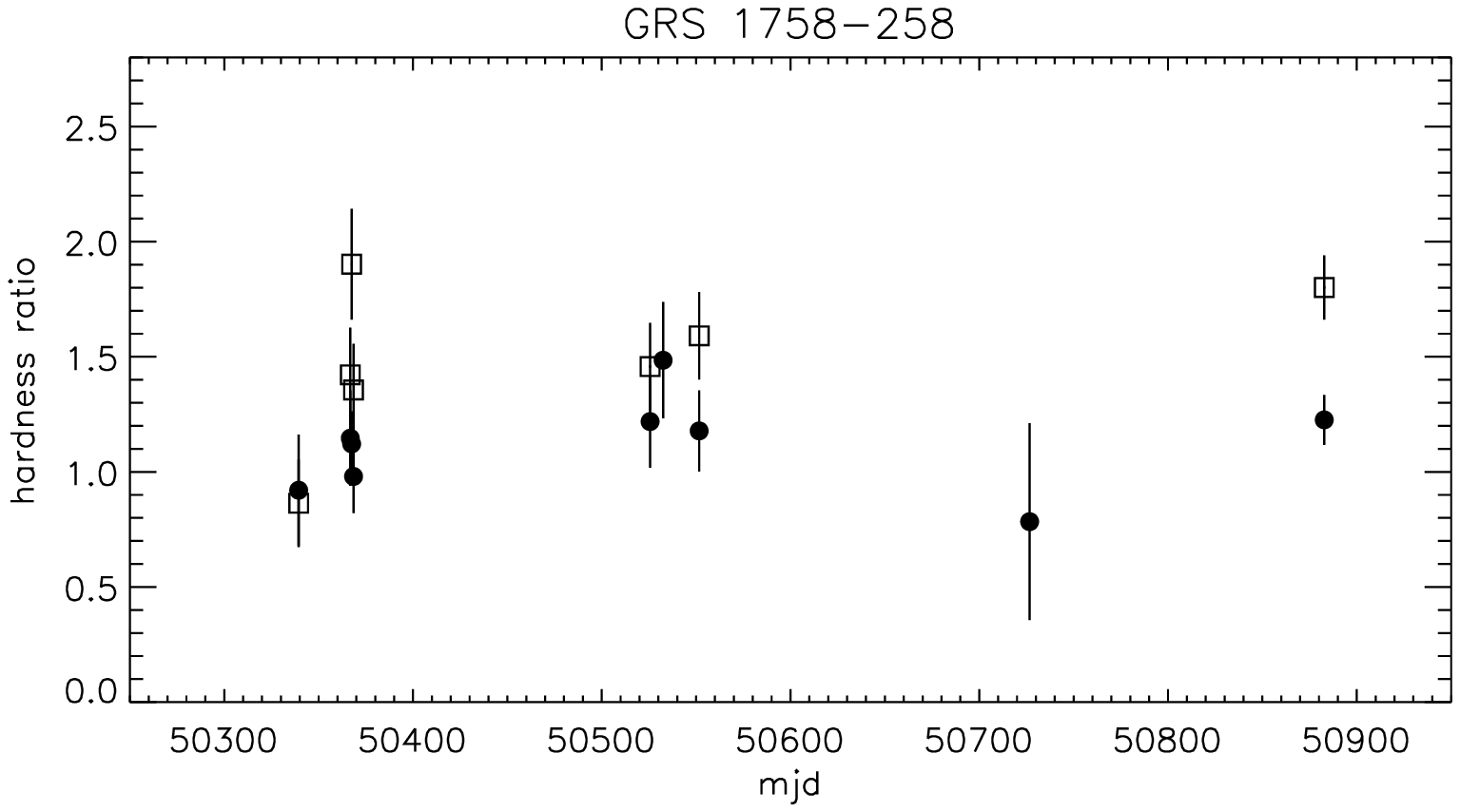, width=11.5cm}}
\caption{FIGURE 2. Time history of the observed hardness ratios. \\
 squares = 10-20 keV / 2-5 keV ratio \\
 dots = 10-20 keV / 5-10 keV ratio
}
\end{figure}
\begin{figure} [hbt]
\centerline{\psfig{file=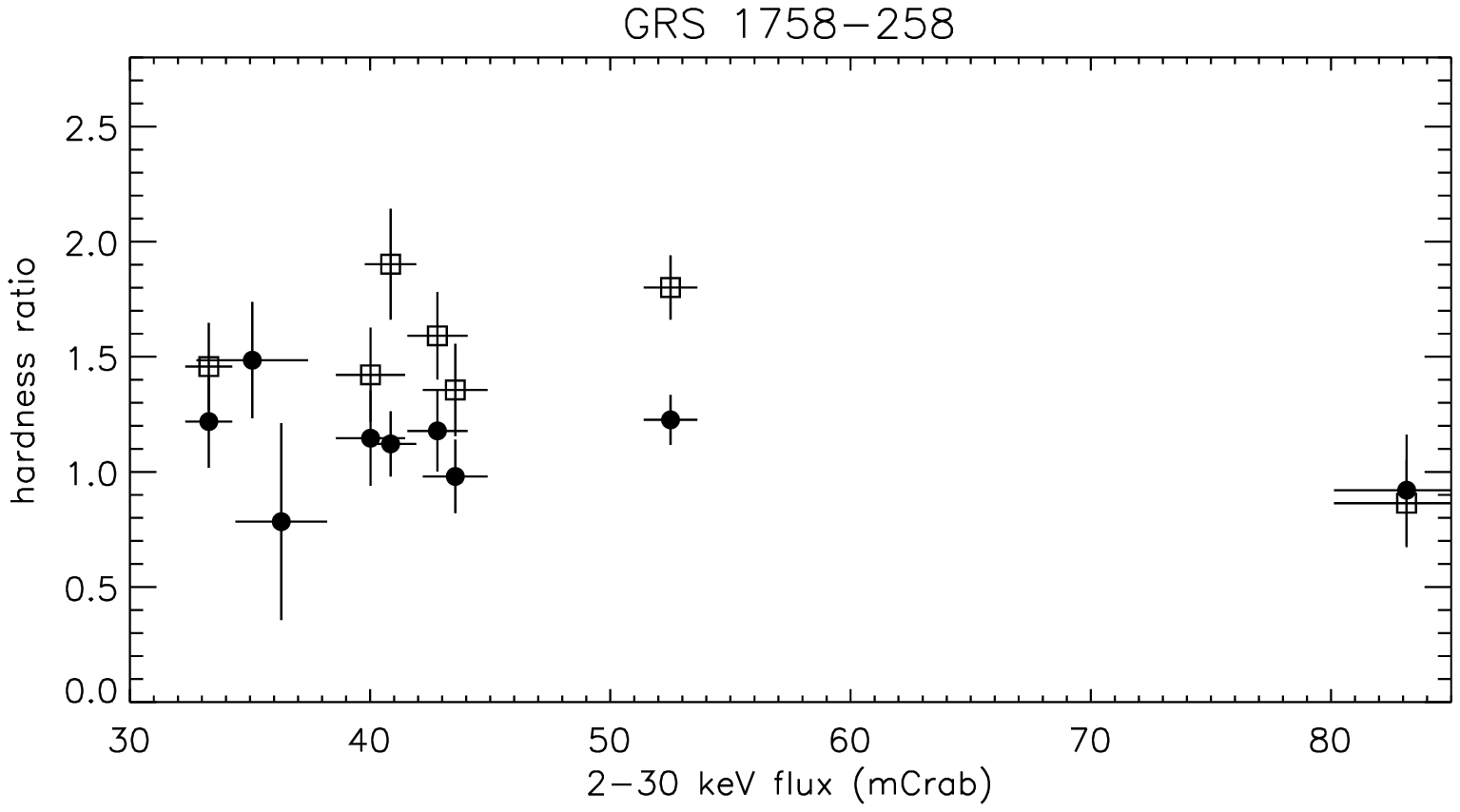, width=11.5cm}}
\caption{FIGURE 3. The observed hardness ratios plotted versus the 2-30 keV source flux. \\
 squares = 10-20 keV / 2-5 keV ratio \\
 dots = 10-20 keV / 5-10 keV ratio
}
\end{figure}

\bsk
\ni 2. OBSERVATION AND DATA ANALYSIS \\
The Wide Field Cameras (WFC) on board the {\em BeppoSAX} satellite, described in detail elsewhere (Jager et al. 1997), consist of 
two identical coded mask telescopes.  The two cameras point at opposite directions each covering 
$40^{\circ} \times 40^{\circ}$ field of view.
The imaging capability, coupled with the good instrument sensitivity (a few mCrab in $10^{4}$ s), allows accurate monitoring of 
complex sky regions like the Galactic Bulge.

The Galactic Bulge region was observed for more than 2 years since August 1996 during a {\em BeppoSAX-WFC} long term 
monitoring campaign (Heise et al. 1997).  About 40 sources, both transient and persistent, have been observed so far
(Ubertini et al. 1998). \\
GRS 1758-258 is in the WFC field of view whenever pointing at this complex sky region and its long term timing and 
spectral behaviour was investigated.  Hereafter preliminary results based on a subset of all the available data are presented.

The observed 2-30 keV source flux on a $\sim600$ days time span is shown in Figure 1.  Thanks to the good angular resolving
power of the cameras ($5^{\prime}$) contamination due to the nearby strong soft source 
GX~5-1, which is $\sim 40^{\prime}$ apart, is negligible.  Our results confirm the variability characteristics of the 
source, since the flux varied by a factor $> 5$ on a $\sim10-30$ days time-scale. The measured average flux ($\sim 35$ mCrab) is 
in good agreement with what was observed by {\em RXTE}/ASM (~25-30 mCrab) at the same epoch (ASM Announcements, http://space.mit.edu/XTE).

Despite the 2-30 keV source variability, our preliminary results do not show strong spectral variability, suggesting the source 
was always observed in the same (hard) state.  In Figure 2 the measured hardness ratios, whenever WFC flux values from the corresponding
energy bands were available, are shown.  The ratios are almost constant throughout the relevant time period (August 1996 - April 1998)
and their values are consistent with a power law spectrum of photon index $-1.7 \pm 0.1$.  \\
Such an hard spectrum is typical of the low state of well established black hole candidates like Cyg~X-1 and it is also regarded 
as the $"$normal$"$ spectral state for the two persistent LMXB microquasars GRS~1758-258 and 1E1740.7-2942 
(Zhang et al. 1997, Grebenev et al. 1997).
There is no clear evidence of a relation between the hardness ratios and the 2-30 keV source flux (Figure 3).

\ssk
\ni 
}

\bsk
\baselineskip = 12pt
{\abstract \ni ACKNOWLEDGMENTS \\
We thank the staff of the BeppoSAX Science Operation Centre and Science
Data Centre for their help in carrying out and processing the WFC Galactic Centre
observations. The BeppoSAX satellite is a joint Italian and Dutch program.
M.C., A.B., L.N. and P.U. thank Agenzia Spaziale Nazionale (ASI) for grant support.
}

\bsk
\baselineskip = 12pt


{\references \ni REFERENCES
\ssk
\ref Gilfanov, M., et al. 1993, ApJ, 418, 844
\ref Grebenev, S.A., Pavlinsky, M.N., and Sunyaev, R.A. 1997, proc. 2nd INTEGRAL Workshop, St. Malo, ESA SP-382, 183
\ref Heise, J., et al. 1997, Nucl. Phys. B (Proc. Suppl.), 69/1-3, in press
\ref Mereghetti, S., et al. 1992, A\&A, 259, 205
\ref Mereghetti, S., Belloni, T., and Goldwurm, A. 1994, ApJ, 433, L21
\ref Mereghetti, S., et al. 1997, ApJ, 476, 829
\ref Mirabel, I.F., et al. 1992, Nature, 358, 215
\ref Sunyaev, R.A., et al. 1991, A\&A, 247, L29
\ref Ubertini, P., et al. 1998, these proceedings
\ref Vargas, M., et al. 1997, proc. 2nd INTEGRAL Workshop "The Transparent Universe", St.Malo, ESA SP-382, 129
\ref Zhang, S.N., Harmon, B.A., and Liang, E.P. 1997, Proc. 4th Compton Symp., AIP 410 (2), 873
}

\end{document}